\documentclass{svproc}
\usepackage{cite}
\usepackage{amsmath,amssymb,amsfonts}
\usepackage{algorithmic}
\usepackage{graphicx}
\usepackage{textcomp}
\usepackage{xcolor}

\usepackage{booktabs}
\usepackage{caption}
\usepackage{multirow}
\usepackage{subcaption}
\usepackage{enumitem}
\usepackage{cleveref}
\usepackage{colortbl}
\usepackage[round]{natbib}

\usepackage{ulem} 
\usepackage{xcolor} 

\begin{document}
\setcounter{secnumdepth}{8}
\mainmatter              
\title{OneLog: Towards End-to-end Software Log Anomaly Detection}
\titlerunning{OneLog}  
%
\author{Shayan Hashemi\inst{1} \and Mika Mäntylä\inst{2}}
\institute{M3S, ITEE, University of Oulu, Finland{1}\\
\email{shayan.hashemi@oulu.fi}\\
Department of Computer Science, University of Helsinki, Finland{2}\\
\email{mika.mantyla@helsinki.fi}
}

\maketitle

\begin{abstract}
With the growth of online services, IoT devices, and DevOps-oriented software development, software log anomaly detection is becoming increasingly important. 
Prior works mainly follow a traditional four-staged architecture (Preprocessor, Parser, Vectorizer, and Classifier). This paper proposes OneLog, which utilizes a single Deep Neural Network (DNN) instead of multiple separate components.
OneLog harnesses Convolutional Neural Network (CNN) at the character level to take digits, numbers, and punctuations, which were removed in prior works, into account alongside the main natural language text. We evaluate our approach in six message- and sequence-based data sets: HDFS, Hadoop, BGL, Thunderbird, Spirit, and Liberty. We experiment with Onelog with single-, multi-, and cross-project setups. 
Onelog offers state-of-the-art performance in our datasets. Onelog can utilize multi-project datasets simultaneously during training, which suggests our model can generalize between datasets. Multi-project training also improves Onelog performance making it ideal when limited training data is available for an individual project. We also found that cross-project anomaly detection is possible with a single project pair (Liberty and Spirit). Analysis of model internals shows that one log has multiple modes of detecting anomalies and that the model learns manually validated parsing rules for the log messages. 
We conclude that character-based CNNs are a promising approach toward end-to-end learning in log anomaly detection. They offer good performance and generalization over multiple datasets. We will make our scripts publicly available upon the acceptance of this paper.

\end{abstract}


\section{Introduction}
\begin{figure*}
  \includegraphics[width=\textwidth]{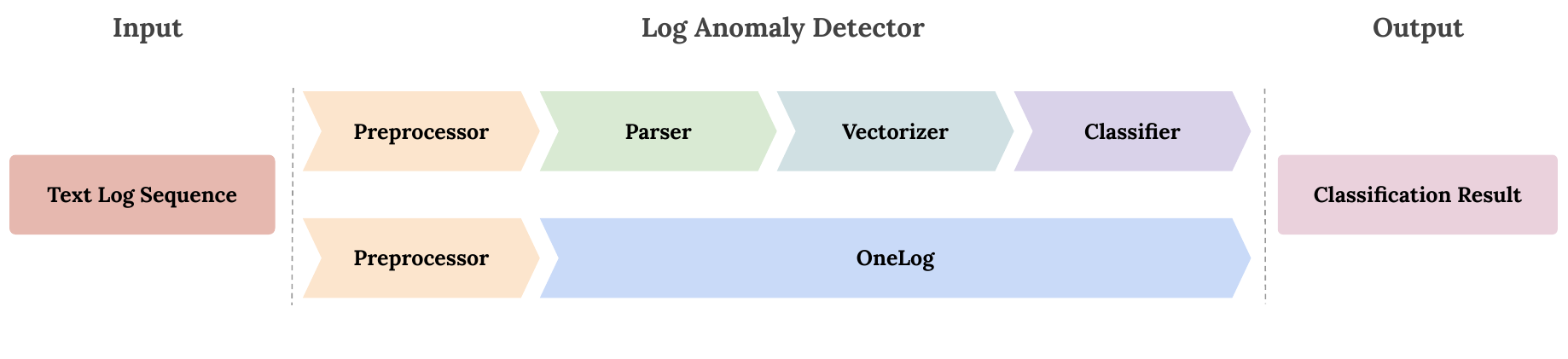}
  \caption{An overall view of the OneLog idea.}
  \label{fig:teaser}
\end{figure*}
Software logs are semi-structured text produced by the software during runtime. A log file includes numerous lines, where each one is comprised of two parts. The structured part, known as header or metadata, consists of information such as timestamp, process ID, thread ID, logging level, and logger component, while the unstructured part, known as message or content, is free from text which developers use to announce the occurrence of an event or report an internal system state.

Recently, deep learning has been used widely to achieve state-of-the-art performance in log anomaly detection \citep{deeplog, logsy, logrobust, loganomaly, cnnlog, sialog}. Most previous works rely on a four-stage architecture (preprocessor, parser, vectorizer, and classifier) to perform their analysis while using deep neural networks only in the final stage (classifier).

In this paper, we propose an alternative where the log parser, vectorizer, and classifier are merged into a single deep neural network, see Figure \ref{fig:teaser}. Merging all components into a single deep neural network takes the log-analysis field one step closer toward end-to-end learning.

Our work is motivated by the potential benefits of end-to-end learning that has demonstrated good accuracy in many different tasks like playing board games \citep{silver2016mastering}, language translation \cite{shen2015minimum}, and autonomous driving \citep{bojarski2016end}.  
End-to-end learning refers to training a single system, usually a deep neural network, to perform the tasks from the raw data to the end results without independent modules or sub-goals. \cite{glasmachers2017limits} explains, this methodology seeks to minimize human intervention by relying solely on data. Potential benefits of end-to-end learning in log anomaly detection are:

\begin{itemize}
    \item Automatic Feature and Task Learning: For instance, while log messaged can be parsed in separate module, the end-to-end approach trusts for the system to autonomously learn log parsing as part of the overarching task, such as anomaly detection, if it proves advantageous. 
    \item Less Engineering Effort: In previous architecture, engineers needed to be experts and optimize three components compared to one. 
    \item Better Accuracy: As explained by \cite{fixesthatfail}, machine-learning systems consist of multiple components that are developed and trained separately may fail to improve the overall performance while each component improves individually. 
    \item Cross-project learning: End-to-end learning makes using data across various projects possible 
\end{itemize}

As our approach is end-to-end, the input is a raw text message or sequence of one or multiple events, while the output is a binary integer indicating sequence anomalousness. Since the input is in raw text, processing could happen at multiple language levels. Nonetheless, this paper proposes a character-based approach rather than a word-based or subword-based one. We chose this because:
\begin{enumerate}
    \item Words take different forms in logs, e.g., the word "request" may be abbreviated to "req" or "block" to "blk"
    \item Processing numerical values is almost impossible in word- and subword-based approaches
    \item Punctuation marks information is preserved
    \item Character-based dictionary is significantly smaller than word-based vocabulary.
\end{enumerate}
Hence, we apply a character-based hierarchical convolutional neural network proposed by \cite{chartextclass} to detect anomalies in software logs.

Moreover, we take it one step further and train multiple datasets together to improve the performance of each individual. Furthermore, we do not stop there and attempt to open the black box of the hierarchical neural network to find an interpretation for our results. In the interest of ensuring transparency, the source code for our experiments has been made publicly accessible via GitHub\footnote{https://github.com/M3SOulu/OneLogReplicationPackage}.

All in all, OneLog's main contributions could be recapitulated within the following research questions:

\begin{enumerate}[label=\textbf{RQ \arabic*.}]
    \item \textit{How accurate is OneLog in software log anomaly detection among datasets of different types (sequence/message based) and domains?} OneLog proves state-of-the-art performance among multiple datasets of different types and domains (see Section \ref{sec:singleproject}).
    \item \textit{Does accuracy improve when combining datasets of the same type and similar domains?} OneLog shows increasing accuracy on the Hadoop dataset when it is combined with the HDFS dataset (see Section \ref{sec:multiproject}). OneLog demonstrates improvements in all datasets when shrunk training sets are combined for training (see Section \ref{sec:multiprojecthard}).
    \item \textit{Is cross-project anomaly detection possible with OneLog?} 
    We cannot be entirely sure about cross-project anomaly detection as we acquired mixed results in our experiments. However, we found very high accuracies when the datasets were similar enough (see Section \ref{sec:supercomp}).
    \item \textit{Is there an interpretation of how OneLog achieves near-perfect state-of-the-art results among all datasets?} Though the black-box nature of neural networks is uninterpretable for humans in the first place, we strived to utilize various methods to obtain insights into our model. We found evidence that internally the neural network learns to operate similarly to previous studies that used separate components to parse, vectorize, and classify log sequences and messages (see Section \ref{sec:interpretation}).
\end{enumerate}

This paper is structured as follows. Next, in Section \ref{sec:met}, we explain our methodology and datasets used in our experiments. Section \ref{sec:res} presents our experimental results in various anomaly detection setups. In Section \ref{sec:interpretation} we continue with experiments and also add explorations on the internal behavior of our model. Section \ref{sec:rel} compares our work against the most relevant related works while Section \ref{sec:lim} presents the limitations of this work. Finally, Section \ref{sec:con} concludes this paper. 

\section{Methodology}
\label{sec:met}
This section explains the idea of OneLog, especially the hierarchical convolutional neural network. Before delving into the model, it is necessary to specify the datasets, their types, and their preprocessing steps.

\subsection{Datasets}
During our study, we faced two different types of datasets. The first is datasets, where each logline is labeled individually. Moreover, these datasets are similar to other one-to-one NLP tasks, such as spam detection or sentiment analysis. These datasets are usually cumulative logs of an extensive system such as operating systems or supercomputers. So, we named this type of dataset Event-based. 

The second type, which we call Sequence-based datasets, are datasets, where a sequence comprised of multiple loglines is labeled instead of individual lines. Additionally, while different sequences may contain loglines of the same event types, each logline could only belong to one sequence. For instance, multiple sequences may begin with loglines of the same format (for example, ``User connected from [ip-address]'') while an individual logline (for example, line 28 for the file ``backend.log'') may only belong to only one sequence. These datasets usually represent processes or behaviors. 

Throughout the rest of this section, we introduce the datasets used in this paper; see Table \ref{tab:datasets}. Later, we will explain using both types in OneLog with no model modification and even training datasets together for a performance boost.

\begin{table*}[t]
\centering
  \caption{Datasets explanation.}
  \label{tab:datasets}
\begin{tabular}{llrrrrrr}
\toprule
\multirow{2}{*}{Dataset} & \multirow{2}{*}{Type} & \multirow{2}{*}{Volume} & \multirow{2}{*}{Log Lines} & \multicolumn{4}{c}{Samples}                    \\
\cmidrule(lr{0.5em}){5-8}
                         & &                                    &                            & Normal     & Anomaly & Total      & Ratio      \\
\midrule
HDFS                     &Sequence & 1.48 GB                            & 11,175,629                 & 558,223    & 16,838  & 575,061    & 0.0301      \\
Hadoop                   &Sequence & 45.7 MB                            & 781,586             & 610        & 15,345  & 15,955     & 25.1557     \\
BGL                      &Event & 708 MB                             & 4,747,963                  & 309,355    & 49,001  & 358,356    & 0.1583      \\
Thunderbird              &Event & 29.6 GB                            & 211,212,192                & 15,512,829 & 24,617  & 15,537,446 & 0.0016      \\
Spirit                   &Event & 37.3 GB                            & 272,298,969                & 13,644,385 & 90,200  & 13,734,585 & 0.0066      \\
Liberty                  &Event & 29.5 GB                            & 265,569,231                & 6,453,814  & 5,114   & 6,458,928  & 0.0008      \\
\bottomrule
\end{tabular}
\end{table*}

\subsubsection{Event-based Datasets}

\paragraph{BGL:} The Blue Gene/L (BGL) dataset \citep{bgl} was gathered from BlueGene/L supercomputer with 131,072 CPUs and 32,768 GB of RAM at Lawrence Livermore National Labs (LLNL) in Livermore, California. The dataset contains a total number of 4,747,963 loglines. However, many samples are duplicated as log events tend to repeat. Removing redundant samples not only reduces the training time and computational cost but also strengthens evaluations' authenticity, as some duplicated samples that may find their way into both the train and test set during the splitting process are preemptively removed. After removing redundancies, the dataset contains 358,356 normal and 49,001 anomalous samples.

\paragraph{Thunderbird:} Containing more than 211,212,192 lines and taking up almost 30 GB of space, Thunderbird dataset \citep{bgl} is one of the most extensive public log datasets. The Thunderbird dataset is collected from the Thunderbird supercomputer system at Sandia National Labs (SNL) located in Albuquerque, with 9,024 processors and 27,072 GB of memory. However, the same as the BGL dataset, Thunderbird includes many duplicated samples. So, to strengthen evaluations, we dismiss the redundant samples. The dataset includes 15,512,829 normal and 24,617 anomalous samples after removing redundancies.

\paragraph{Liberty:} The Liberty supercomputer, with 512 processors and 944 GB of memory, is located at Sandia National Labs (SNL) in Albuquerque. The dataset \citep{bgl} contains 265,569,231 lines, taking almost 30 GB of space. Not being an exception from others, the dataset is not divine from duplication. After removing redundancies, the dataset remained with 6,453,814 normal and 5,114 anomalous samples.

\paragraph{Spirit:} The Spirit supercomputer at Sandia National Labs (SNL) located in Albuquerque is equipped with 1028 processors and 1024 GB of memory. Although it was ranked 202 among the best supercomputers in the world between 2004 and 2006, it is relatively weak compared to today's standard. The Spirit dataset \citep{bgl} is collected from the supercomputer containing 272,298,969 lines of logs taking up 37 GB of space. After redundancy removal, however, the dataset contains 90,200 normal and 13,644,385 anomalous samples.

\subsubsection{Sequence-based Datasets}

\paragraph{HDFS:}
Hadoop Distributed File System (HDFS) is a fault-tolerant and low-cost distributed file system established by Hadoop. The dataset is regarded as a benchmark in the log anomaly detection domain. It was constructed via map-reduce jobs with more than 200 Amazon EC2 nodes, and it was annotated by Hadoop domain specialists \citep{hdfs}. Each log message's $block id$ has been used to produce sequences. We retrieve 16,838 anomaly and 558,223 non-anomaly samples after the preprocessing.

\paragraph{Hadoop:} Hadoop is a big data processing technology that enables distributed data processing. The dataset, introduced by \cite{hadoopdataset}, comprises a five-machine Hadoop cluster log, each with an Intel(R) Core(TM) i7-3770 CPU and 16GB of RAM. Furthermore, since our algorithm only identifies anomalies, not their categories, we first combine all types of anomalies into a single class. Then, as many sequences in the dataset are excessively lengthy, we use the sliding window approach (window size of 64) to produce subsequences. As the merging resulted in a higher number of anomaly samples, we inverse the labels (replace zeros with ones and ones with zeros) to maintain the nature of the anomaly detection task where the minority is labeled as positive. Thereafter, the dataset includes 610 anomaly and 15,345 non-anomaly samples. On the other hand, when the dataset couples with other datasets in training, we do not inverse the labels as other datasets compensate for the inversed imbalances of Hadoop.

\subsection{Preprocessing}
Since the Hierarchical CNN model processes input data in sequence form (a matrix of characters where each row represents an event, so the entire matrix represents a sequence of messages), log input should be transformed into sequences before they are passed to the model, similar to the work by \cite{deeplog}. As mentioned previously, in Table \ref{tab:datasets}, HDFS and Hadoop's sequences are produced by a particular property, naming $block_id$ in HDFS and files in Hadoop. However, BGL, Thunderbird, Spirit, and Liberty are event-based datasets which makes sequences meaningless for them. So, we produce single-event sequences (Each event is made into a sequence that contains only one event). After sequence production, environmental biases in the dataset, such as IP addresses and integer sections of block IDs, are removed to avoid creating any bias in the model. Then, each sequence is converted to a matrix of integer numbers according to a character table.

\subsection{Hierarchical CNN Model}
As previously stated, OneLog treats the log anomaly detection problem as a sequence of the text classification problem. Moreover, it takes its input in raw text format (a matrix of characters) and learns to classify anomaly sequences during the training. Thus, since the characters are encoded into integer numbers, the entire task is to classify a matrix of integers.

\begin{figure}[tb]
  \centering
  \includegraphics[width=\textwidth]{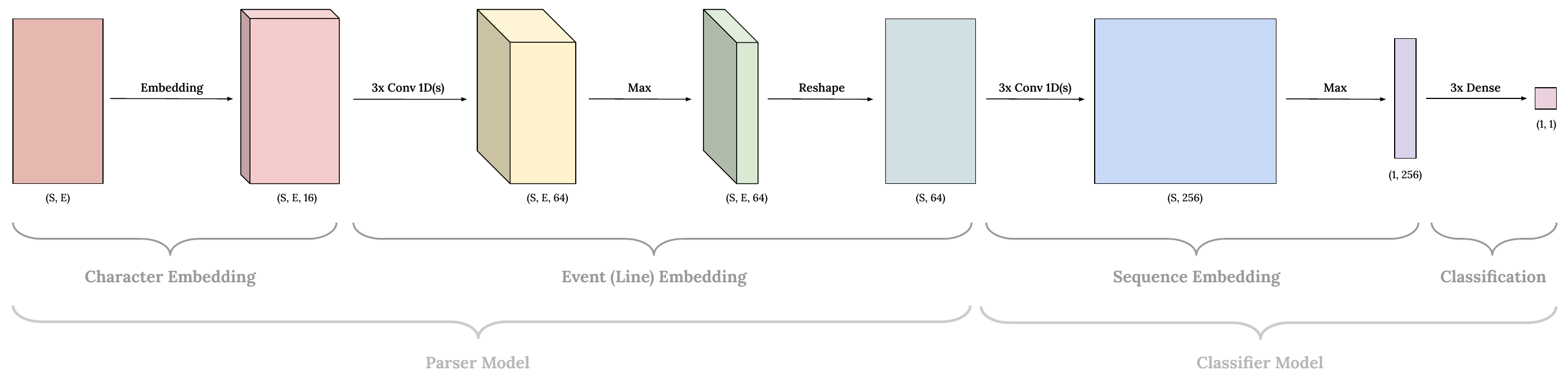}
  \caption{The Hierarchical CNN model visualization. Variables $S$ and $E$ are indications of sequence and event lengths which are equivalents of $L_s$ and $L_l$ in the model explanation text. However, It must be noted that within our experiments, though the sequence events window size is 64, there is no limit to the number of characters in each log event (line). }
  \label{fig:onelogarch}
\end{figure}

Accordingly, the model is divided into three sections in our network. The first one embeds input characters, the second processes embedded characters to form event vectors, and the third processes the embedded vectors and classifies the entry (the character matrix), see Figure \ref{fig:onelogarch}. 

In more detail, the input is a sequence of log events, while each log event is a sequence of characters (the leftmost rectangle in the figure). Hence, assuming that $B$ is the batch size used in our model training, $L_s$ is the maximum number of log events in a sequence, and $L_l$ is the maximum number of characters in a log event (line), the input size is $(B, L_s, L_l)$. At first, OneLog embeds all characters into an arbitrary vector space with the dimension of $C_e$, the character embedding matrix's dimensions, producing an intermediate tensor shape of $(B, L_s, L_l, C_e)$. After that, three one-dimensional convolutional layers (event embedding network) are applied to each log event (the second dimension) in the tensor. Presuming that $C_l$ is the number of output channels in the log event embedding network, the intermediate tensor shape is $(B, L_s, L_l, C_l)$. The output is aggregated by taking the maximums in the third dimension, delivering the intermediate tensor shape of $(B, L_s, C_l)$.

Following the aggregation, the intermediate tensor goes through three one-dimensional convolutional layers (sequence embedding network). Supposing the number of output channels in the sequence embedding network's last layer is $C_s$, the intermediate tensor shape is $(B, L_s, C_s)$. Next, the output is aggregated by selecting the maximums in the second dimension, making a tensor shape of $(B, C_s)$. Finally, the intermediate tensor passes through multiple fully-connected layers activated by a Rectifier Linear Unit (ReLU) followed by a Sigmoid-activated one with one output neuron, producing an output shape of $(B, 1)$, which indicates the probabilities of input sequences being anomalous. As the model hierarchically applies convolution, we call it Hierarchical Convolutional Neural Network (HCNN). 

Since the proposed model is entirely made of convolutional and fully connected layers, the computations may be parallelized, making the training and inference drastically more efficient than RNN-based solutions. Furthermore, parallelism allows for higher quantities of training data that boost performance and generalizability.

\section{Experimental Results}
\label{sec:res}

Throughout this section, OneLog is assessed in different challenges, from standard ones to more challenging ones, using the aforenamed datasets. More in-depth, we evaluate OneLog in a benchmark challenge that trains and tests the model on the same data. To attain near-perfect results for those that did not achieve it in the first experiment, we combine correlated datasets to increase their performance in the second experiment. Following that, we evaluate the effects of multi-project training by reducing the training set samples from all datasets. Finally, we evaluate the idea of cross-project training in supercomputer log datasets.

\subsection{Environment Settings}
It is important to discuss some experiment environment details before diving deep into the experiments. Likewise, the HierarchicalCNN model is part of the deep neural networks family, which usually requires a considerable amount of data to generalize appropriately. On the other hand, a decent number of anomalous samples is mandatory in the test set to perform a consolidated evaluation. So, we decide to split each dataset 80\% to 20\% between train and test sets for the performance evaluation experiments. We believe an 80-20 split is a good sweet spot between training comprehensivity and testing rigidity. Furthermore, we preserve 20\% of the training data for validation purposes to prevent overfitting. Thus, about 74\% of the entire dataset is used for training.

On the other hand, since we train our model on lots of data, we require a fast method of training to make the training method viable. Hence, we use mixed floating point precision during our training, maintaining the model in 32-bit floating point numbers while using 16-bit floating point numbers for gradient calculation. The mixed floating point precision method allows for larger training batches possible. However, batch sizes differ for each dataset as it is correlated to the average sequence size, which is different for each dataset. So, we used smaller batches in training on datasets with larger average sequence lengths, like Hadoop, and larger ones on datasets with shorter average sequence lengths, such as Spirit and Thunderbird. 

Furthermore, as the HierarchicalCNN model contains multiple convolutions and fully-connected layers (nine layers in total), it is drastically prone to the vanishing gradient problem. In order to circumvent this problem, we utilize the batch normalization technique in all convolution layers of message and sequence embedding sections. Doing so facilitates the flow of gradient throw the network and expedites the convergence process.

Since we are utilizing mixed floating point precision to expedite training speed, our model is exposed to numerical instability during the training process. The instabilities come from computing the derivative of the natural logarithm and exponential in a mixed precision fashion. To compensate for this problem, instead of using the traditional binary classification loss (a variety of log loss) on a probability output, we use the binary classification loss for logits and remove the sigmoid activation at the end of the model, making the last layer of the model linear. Thus, we have to compute neither the exponential function within the sigmoid nor the natural logarithms within the log loss, enhancing the numerical stability of our training. Furthermore, we use the Adam optimizer to produce a more robust gradient value.

\subsection{Single-project experiment - RQ1}
\label{sec:singleproject}
\textbf{RQ1: }How accurate is OneLog in software log anomaly detection among datasets of different types (sequence/message based) and domains?

\subsubsection{Motivation} In this experiment, we strive to evaluate the HierarchicalCNN model among different datasets and compare them to parser-less and parser-based state-of-the-art log anomaly detection methods.

\subsubsection{Method} In order to perform the evaluation process, we need to train the model on each dataset. However, before training, we acquire the optimum batch size for each dataset. The optimum batch size is the largest batch size that could fit in the accelerator's (GPU in our experiment) memory and perform all necessary training steps (feedforward, backward gradient computation, and weight update). A larger batch size compared to a smaller one not only reduce the training time due to better parallelization but also results in more generalized gradient values that help with faster convergence. 

It is worth mentioning that since the batch sizes were not deterministic values and were entirely dependent on the machine's hardware specifications and state, we did not include them in the results. However, we can confirm that since Hadoop's average sequence length is larger, smaller batch sizes were consistently found for it compared to single-event sequence datasets such as BGL or Thunderbird. 

Furthermore, since Hadoop does not include enough samples, it puts the model at risk of overfitting. So, to prevent overfitting, we use dropout layers with the probability of 0.5 for all convolution and fully-connected layers in the HierarchicalCNN model while experimenting on the Hadoop dataset.

Thereafter, we train each model with the optimum batch sizes that were previously found. Furthermore, training for each model happened no more than 128 epochs to further standardize the evaluation. Finally, we evaluated each trained model using each dataset's test set by computing the $F_1$ score for each test set and comparing it to the state-of-the-art methods.

\subsubsection{Results} 
As Table \ref{tab:singleresults} shows, OneLog achieves near-perfect results, either close to or better than the state-of-the-art. OneLog scores the $F_1$ score of 0.99 on all datasets except for Hadoop. We think the reason for such an incident is that Hadoop is a more complex dataset with fewer samples, making it extremely difficult for the model to score as high in this dataset. We provide a solution to improve the model's performance on the Hadoop dataset in the next experiment.

\begin{table*}[t]
\centering
  \caption{Evaluation of the OneLog trained on the multi-project dataset and its comparison to the state-of-the-art (to the best of our knowledge) method on each dataset.\\{\small \textbf{Note:} In methods denoted by a superscript asterisk, we were unable to replicate the results with equivalent accuracy, thus we have cited the findings from the original publication.}}
  \label{tab:singleresults}
\begin{tabular}{lrrrrrr}
\toprule
Method    & HDFS & Hadoop & BGL  & Thunderbird & Spirit & Liberty \\
\midrule
OneLog    & 0.99 & 0.97   & 0.99 & 0.99        & 0.99   & 0.99    \\
LogBERT\text{*} \citep{logbert}   & 0.82 & -      & 0.91 & 0.97        & -      & -       \\
NeuralLog \citep{anodetnoparsing} & 0.98 & -      & 0.98 & 0.96        & 0.97   & -       \\
Logsy\text{*} \citep{logsy}    & -    & -      & 0.65 & 0.99        & 0.99   & -       \\
LogRobust\citep{logrobust}  & 0.99 & 0.90   & 0.75 & 0.68        & 0.95   & -       \\
LogAnomaly \citep{loganomaly}   & 0.94 & -      & 0.88 & 0.84        & 0.95   & -       \\
DeepLog\citep{deeplog}    & 0.95 & -      & 0.86 & 0.93        & -      & -       \\
SiaLog\citep{sialog}    & 0.99 & 0.94   & 0.99 & -           & -      & -       \\
CNNLog\citep{cnnlog}     & 0.98 & 0.92   & 0.95 & -           & -      & -       \\
Auto-LSTM \citep{autoblstm} & -    & -      & 0.95 & 0.99        & -      & -      \\
\bottomrule
\end{tabular}
\end{table*}


\subsubsection{Discussion}
We think the excellent performance of OneLog could be attributed to three facts. First, even past works have achieved relatively high $F_1$ scores in evaluation datasets, suggesting that achieving high performance with multiple approaches in that dataset is possible. Second, our character-based NLP approach demands samples on a large scale and uses them to train millions of parameters to learn complex behavior in the data. Thirdly, the Hierarchical CNN, operating at the character level, can meticulously focus on parameters. In specific scenarios, these parameters crucially determine a sequence's anomaly label.

\begin{figure}[t]
  \centering
  \includegraphics[width=\columnwidth]{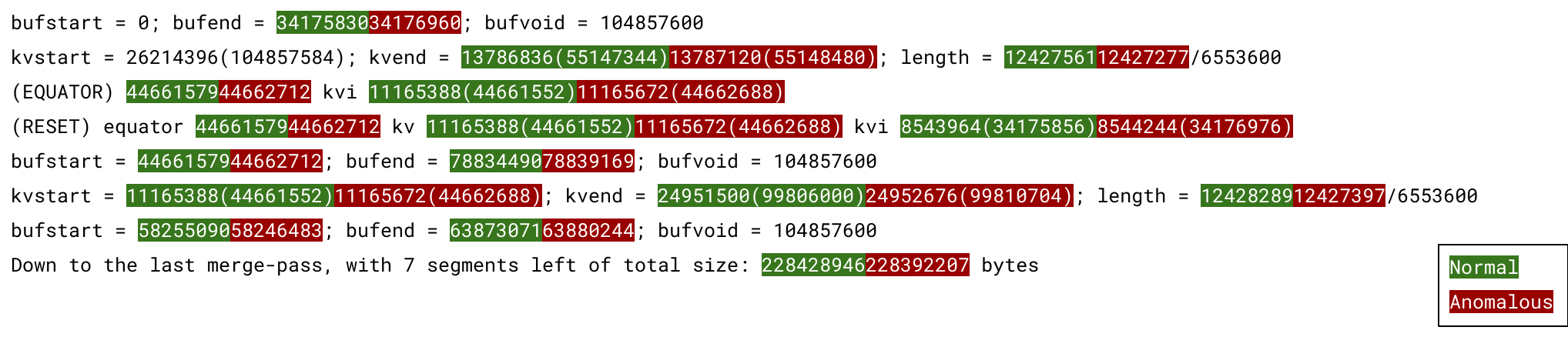}
  \caption{A sample from the Hadoop dataset where parameters are the sole determinants differentiating between normal and anomalous behaviors.}
  \label{fig:hadoop_diff}
\end{figure}

\begin{figure}[t]
  \centering
  \includegraphics[width=\columnwidth]{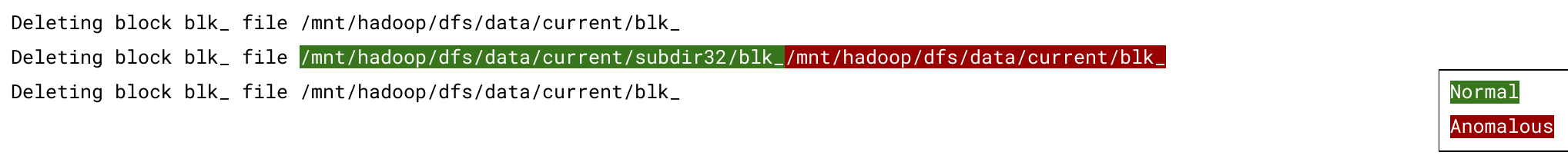}
  \caption{A sample from the HDFS dataset where parameters are the sole determinants differentiating between normal and anomalous behaviors.}
  \label{fig:hdfs_diff}
\end{figure}

For instance, Figure \ref{fig:hadoop_diff} and \ref{fig:hdfs_diff} show two examples where parameters play a significant role in anomaly detection. In the HDFS dataset Figure  \ref{fig:hdfs_diff}, while two sequences with the \texttt{block\_id}s \texttt{blk\_-1877795964140566716} and \texttt{blk\_8557779918081539564} follow the same execution path, they have distinct labels, with their parameters being the distinguishing factor. Specifically, \texttt{blk\_-1877795964140566716} removes three block files from \texttt{/mnt/hadoop/dfs/data/current/}. However, the standard procedure is to remove the second file from \texttt{/mnt/hadoop/dfs/data/current/subdir}, precisely what \texttt{blk\_8557779918081539564}  accomplishes. Additionally, in the Hadoop dataset Figure \ref{fig:hadoop_diff}, two sequences from containers \texttt{container\_1445087491445\_0007\_01\_000008} and \texttt{container\_1445094324383\_0001\_01\_000009} differ only in their parameters. The former is categorized as normal, while the latter is identified as a machine-down anomaly. Such detailed distinctions are feasible only with accurate parameter processing, a strength inherent in the Hierarchical CNN model.


On the other hand, we believe the model performed less accurately on the Hadoops dataset since it does not contain enough samples. The upcoming experiment confirms this hypothesis to some degree.

\subsection{Multi-project experiment for sequence-based datasets - RQ2}
\label{sec:multiproject}
\textbf{RQ2:} Does accuracy improve when combining datasets of the same type and similar domains?
\begin{table}[t]
\centering
  \caption{Evaluation of the Multi-project (MP) OneLog trained on the multi-project dataset and its comparison to the state-of-the-art (to the best of our knowledge) method on each dataset.}
  \label{tab:multiresults}
\begin{tabular}{lr}
\toprule
Method                  & Hadoop \\
\midrule
OneLog (Hadoop only)    & 0.97   \\
OneLog (Hadoop \& HDFS) & 0.99   \\
LogRobust               & 0.90   \\
SiaLog                  & 0.94   \\
CNNLog                  & 0.92  \\
\bottomrule
\end{tabular}
\end{table}

\subsubsection{Motivation} As shown in the previous experiment, OneLog achieved near-perfect results in all datasets except the Hadoop dataset. We think the root cause of this is Hadoop's lack of samples (610 normal and 15,345 anomalies) and overcomplexities compared to other datasets. So, a potential solution could be transferring knowledge from other datasets. After all, since we use a character-based approach, it may capitalize on the shared vocabulary among different log datasets. Therefore, in this experiment, we combine multiple datasets for training.

\subsubsection{Method} We train our HiearchicalCNN model on a combined dataset of multiple log datasets. Although it seems straightforward, choosing suitable datasets could be challenging. On the one hand, the more data fed to the model, the more generalized the model becomes. On the other hand, however, our arsenal of datasets is extensive, making it computationally expensive for training. Thus, we explored a handful of datasets that increase the Hadoop results to near perfect while maintaining trainability. After attempting multiple trainable combinations, we found that combining Hadoop with HDFS increases the model's performance on Hadoop to near-perfect while keeping the model trainable.

\subsubsection{Results} As shown in Table \ref{tab:multiresults}, with the help of the HDFS dataset, Multi-project OneLog achieves near-perfect results, $F_1$ score of 0.99, compared to the Single-project OneLog that had the $F_1$ score 0.97 on the Hadoop dataset.

\subsubsection{Discussion}
Although the extra performance may not be worth the computational costs in some cases, this experiment bears witness to the fact that combining multiple log datasets improves OneLog's performance on an individual dataset. This approach is handy when the available data is insufficient to train a model with desired performance. In such cases, we may take advantage of other datasets (such as open public ones) to boost the performance. Furthermore, this finding is crucial as it shows that having a generalized model for software log anomaly detection is possible.

\subsection{More difficult and comprehensive multi-project experiment - RQ2}
\label{sec:multiprojecthard}
\begin{table}[t]
\centering
  \caption{Multi-project training when all datasets are equally difficult.}
  \label{tab:eqmult}
\begin{tabular}{lrrr}
\toprule
Dataset     & Subset Ratio & Single-project & Multi-project \\
\midrule
HDFS        & 0.0090                & 0.83               & 0.86             \\
Hadoop      & 0.7800                & 0.78               & 0.89             \\
BGL         & 0.0150                & 0.75               & 0.98             \\
Thunderbird & 0.0025                & 0.74               & 0.91             \\
Spirit      & 0.0020                & 0.80               & 0.97             \\
Liberty     & 0.0100                & 0.77               & 0.89            \\
\bottomrule
\end{tabular}
\end{table}

\textbf{RQ2:} Does accuracy improve when combining datasets of the same type and similar domains?

\subsubsection{Motivation} As shown in Section \ref{sec:multiproject}, combining multiple datasets increases OneLog's performance on the Hadoop dataset. However, it was impossible to observe any performance improvement in other datasets as the results were already near-perfect  ($F_1$ score of 0.99) in single project experiments; see Table \ref{tab:singleresults}. 

Given a large amount of data, the task was too easy for our deep neural network. However, having more than sufficient data is a luxury that might not hold true in an industry where things like time pressure hampers practitioners, and the systems are constantly evolving. Therefore, we decided to reduce the training data to make the anomaly detection task more difficult. This allows us to experiment more on the Multi-project Onelog performance and assess if multiple datasets help the model perform better at each dataset.

\subsubsection{Method} Reducing training set size is challenging as the datasets at hand vary in nature, complexity, and volume. For instance, the HDFS dataset contains many similar sequences and not many templates, so it is on the easier end of the spectrum. However, for the Hadoop dataset, since the data volume (number of sequences) is lower and the complexity (number of templates) is higher, the model finds it more challenging to achieve near-perfect results.

At first, we decided to take a constant percentage of each dataset. However, we found this method inapplicable as the trained model's accuracy varied greatly between datasets. Furthermore, the model achieved the same near-perfect result just by utilizing 2\% of the HDFS training set. On the other hand, the model misclassified nearly all sequences when it was trained on 10\% of the Hadoop training set.

Thus, we changed the dataset reduction method from a constant percentage to the same difficulty for all. Going more in-depth, we started from a complete training set and decreased the available training set ratio until the accuracy had gotten to a certain level in single-project training and repeated this process for each dataset individually. We aimed to get an $F_1$ score of 0.8 for all datasets as it allows some learning to happen in each single-project training while leaving enough room for improvement. One can notice from Table \ref{tab:eqmult} that not all datasets got to 0.8 in a single-project setup, as the reduction in training data resulted in relatively abrupt performance changes. Nevertheless, by reducing the training set samples, we achieved the goal of reducing single-project performance while allowing some learning to appear. 

After reducing the training data, we combine all subsets and train a Multi-project Onelog model using the combined dataset. We measure the performance of the trained model on each test set individually. Finally, it is important to mention that we only reduced training set sizes and left test sets untouched in this experiment, implying that they are the same as in the previous experiments.

\subsubsection{Results} 
Table \ref{tab:eqmult} shows how the model performs on each test set after being trained in single-project and multi-project settings. We get improvement in all datasets. The average single-project $F_1$ score is 0.78, while the average multi-project $F_1$ score is 0.92. The biggest improvement is in BGL, which improves the test set $F_1$ score from 0.75 to 0.98 in single-project and multi-project training, respectively. The lowest improvement is HDFS which improves only from 0.83 to 0.86 in single-project and multi-project training, respectively. 

\subsubsection{Discussion}
The significant average improvements and the fact that we got improvements in all datasets provide evidence that multi-project training improves performance. The results suggest that the OneLog model can generalize on multiple datasets and that this generalization results in performance improvement when sufficient training data is unavailable.

\subsection{Cross-project experiment with supercomputer log datasets - RQ3}
\label{sec:supercomp}
\textbf{RQ:} Is cross-project anomaly detection possible with OneLog?
\subsubsection{Motivation}
As demonstrated in Sections \ref{sec:multiproject} and \ref{sec:multiprojecthard}, combining multiple datasets improves performance. However, there might exist situations when no training data is available due to data inaccessibility, incompatibility, or scarcity. In such situations, cross-project training may become the only viable option. Assuming we have multiple similar datasets, in cross-project training, a model trains on one or multiple datasets while being tested on other datasets that it has not seen in training. Such a method allows software engineers to train a model once and use it as the system changes and also use a trained model in situations where they do not have access to the data.

\subsubsection{Method}
Before diving deep into the experiments, we must select the experimentation datasets. We select BGL, Thunderbird, Spirit, and Liberty as they are all supercomputer logs and, thus, possess a reasonable chance that cross-project training would offer positive results. In order to evaluate the cross-project training idea, we developed two experiments. During the first experiment, we train a distinct model for each dataset and evaluate its performance on all other datasets. This way, we evaluate if we could deploy a model to detect supercomputer anomalies without being trained on the same data. In other words, the models are the same as Section \ref{sec:singleproject}; however, evaluation datasets are other supercomputer logs than the training dataset.
In the second experiment, we leave one of the datasets out for evaluation while training on the rest. This experiment examines the possibility of training a model on similar datasets and deploying it to another machine if the deployment machine's log were inaccessible.

\begin{table}[t]
\centering
\caption{Cross-project training results within train on one evaluate on others settings.}
\label{tab:scc}
\begin{tabular}{lrrrr}
\toprule
\multirow{2}{*}{Training} & \multicolumn{4}{c}{Evaluation}        \\
\cmidrule{2-5}
                          & BGL  & Thunderbird & Spirit & Liberty \\
\midrule
 { BGL}             & 0.99 & 0.00        & 0.00   & 0.00    \\
 { Thunderbird}     & 0.00 & 0.99        & 0.10   & 0.01    \\
 { Spirit}          & 0.00 & 0.06        & 0.99   & 0.91    \\
 { Liberty}         & 0.00 & 0.05        & 0.99   & 0.99    \\
\bottomrule
\end{tabular}
\end{table}

\begin{table}[t]
\centering
\caption{Cross-project training results within leaving one out for evaluation settings.}
\label{tab:cross}
\begin{tabular}{llllr}
\toprule
\multicolumn{4}{c}{Datasets}                      & \multirow{2}{*}{ $F_1$ score} \\
\cmidrule{1-4}
\multicolumn{3}{c}{Training}        & Evaluation  &                             \\
\midrule
Thunderbird & Spirit      & Liberty & BGL         & 0.00                        \\
BGL         & Spirit      & Liberty & Thunderbird & 0.05                        \\
BGL         & Thunderbird & Liberty & Spirit      & 0.99                        \\
BGL         & Thunderbird & Spirit  & Liberty     & 0.99                        \\
\bottomrule
\end{tabular}
\end{table}

\subsubsection{Results}
As Table \ref{tab:scc} illustrates the first experiment's results, cross-project training could be possible under the condition that datasets are similar enough. Training the model on Spirit data produced an $F_1$ score of 0.91 in Liberty. Vice versa, we observed that training on Liberty results in an F1-Score of 0.99 in Spirit. However, all other pairs offered poor results, e.g., training on Thunderbird produces $F_1$ scores of 0.10, 0.01, and 0.00 in Spirit, Liberty, and BGL, respectively. 

Table \ref{tab:cross} discloses the second experiment's results, which are in line with the previous one. For Spirit and Liberty, we got an $F_1$ score of 0.99 when the model was trained on the three other supercomputers' datasets. The result indicates that the model may not be trained on the target machine's data at all should a minimum of one similar dataset exist within the training datasets pool. 

\subsubsection{Discussion}
It appears that we could achieve high accuracy even by training on foreign data only. However, this foreign data has to be close enough as we could find only one pair (Spirit and Liberty) where cross-project training produced accurate enough results. An important question for future research are methods of finding close enough datasets before training.

\section{Model Interpretation - RQ4}
\label{sec:interpretation}
\textbf{RQ4:} Is there an interpretation of how OneLog achieves near-perfect state-of-the-art results among all datasets?

In the previous section, we saw how OneLog outperforms the state-of-the-art methods among different datasets. Although one may assert that the model internals do not matter as long as the performance is good, we believe interpreting the black-box model of deep learning increases authenticity. More in-depth, we can find out if models have found an unexpected method in their decision-making process, such as making decisions based on an environmental bias in the datasets (such as IP address). 
Nonetheless, OneLog's deep neural network's black box remains a mystery, as it is not apparent how it achieves such high scores. This section aims to comprehend the deep neural network and interpret its decision-making.

Although some algorithms have been proposed to interpret deep neural networks' decision-making \citep{gradstar,shapley,integratedgrad, layerprop, deeplift}, especially in the computer vision field, we found their outcomes impractical for our study during our experiments. As the mentioned methods discover the most contributing input(s) to the production of the output while our model's input is a matrix of characters, we found the idea of the most contributing character(s) to a log sequence anomalousness irrational.

Furthermore, we consider that our deep neural network's hierarchical architecture makes it unintuitive to interpret it as one neural network, as different model sections are supposed to perform different tasks. Furthermore, we believe each model section should be interpreted separately. Hence, we split our network into two parts and investigated them separately.

Accordingly, we split the hierarchical convolutional neural network into two submodels. The first submodel is comprised of character and event embedding layers (see Figure \ref{fig:onelogarch}), while the second submodel carries sequence embedding and classification layers. We name the first submodel the "Parser Model" and the second submodel the "Classifier Model".

Throughout the rest of this section, we strive to investigate each model's internal decision-making mechanisms and evaluate how accurately they are performing the task considered for them.

\subsection{Parser Model Interpretation}
\subsubsection{Background}
This subsection is dedicated to figuring out the under-the-hood mechanics of the Parser Model. Accordingly, we intend to know if the Parser Model acts similarly to log parsers in prior studies by \cite{loghubtools}. Although neural networks and standard parsers are both functions, a direct comparison is impossible. Furthermore, a neural network (specifically the Parser Model) is a sequence of linear-algebra operations on input vector(s), while parsers are considered as a set of templates that are compared against the input to find the fittest template. Therefore, we found these two models incomparable. 
So, a direct function comparison between a neural network and a standard parser is impossible. 

As an internal comparison is impossible, function similarity approximation becomes a viable option. Moreover,  the similarity between two functions within a defined domain could be approximated by computing the difference between their outputs for a set of distributed inputs from the desired domain. So, we may compute the similarity between the Parser Model and an actual log parser by comparing their outputs for a specific set of inputs. Nonetheless, this raises another challenge as the Parser Model's output is a continuous vector while a parser's output is a discrete categorical cluster number.

The foremost solution to this problem may be clustering vectors and comparing the clusters with the ground truth categories using a standard clustering performance metric. However, since a perfect clustering algorithm does not exist, by doing so, we are including clustering errors in our measurements. Thus, to circumvent the latest trouble, we use silhouette score \citep{silho,silho2}. Though silhouette score is not commonly employed for this goal, it fits our situation flawlessly.

\subsubsection{Method}
The silhouette score is an unsupervised metric for clustering accuracy measurement. Silhouette ranges from -1 (worst) to +1 (best). According to \cite{scikit-learn}, the score is 1 when all clusters are correct, it is -1 when all clusters are incorrect, and zero indicates overlapping clusters. Silhouette indicates how close the members of each cluster are and how distant they are from other clusters simultaneously. Accordingly, lower intra-cluster and higher extra-cluster distances elevate silhouette scores. So, if the silhouette score is passed with embedded vectors and the ground truth template numbers (manually labeled), it shows how close events of the same template are embedded while being distant from templates of other clusters. This measures how close the Parser Model's output is to manually-labeled event templates that act as ground truth. Moreover, it answers the question \textit{Is Parser Model embedding events according to their templates, which could be deemed as an act of log parsing?} If the answer to this question is yes (positive silhouette score), then it may be concluded that the Parser Model is actually parsing the log line before passing them to Sequence Embedding layers.

We acquire the gold standard human-labeled log event template datasets for our ground truth. Past work has used this data for log parser performance experiments \citep{loghubtools}. Since the parsed template data is unavailable for all of our test datasets, we performed this experiment for only four datasets (HDFS, Hadoop, BGL, and Thunderbird). During the experiment, for each dataset, we first embedded all events from the log parser benchmarking dataset using a Parser Model extracted from a model trained on the target dataset and calculated the silhouette score for embedded events using their template number.

In addition to the silhouette score, we provide a visualization. Visualization of more than three dimensions is uninterpretable for humans. Thus, we reduce embedded event vector dimensions using the U-MAP algorithm \cite{umap} and visualize the events in a 2D plot. Each event is also colored based on its template. 

\subsubsection{Results}
Table \ref{tab:silouette} shows the silhouette scores for different datasets.  From the table, we can observe that we achieve an excellent score of 0.74 for Hadoop. For HDFS and BGL, on the other hand, the scores are lower yet still clearly on the positive side, with silhouette scores of 0.34 and 0.24, respectively, while Thunderbird stands in the middle ground between BGL, HDFS, and Hadoop with silhouette scores of 0.48. So it appears that the Parser Model is, in fact, parsing the events in a similar way as other log parsers. We should bear in mind that our Parser Model is not trained for log parsing, and all the correct event labeling it achieves is simply a side product of its main goal of anomaly detection. 

\begin{table*}[tb]
    \centering
  \caption{The silhouette score of the Parser Model with respect to manually labeled log templates.}
  \label{tab:silouette}
    \begin{tabular}{lrrr}
    \toprule
    Dataset     & Samples & Templates & Silhouette Score    \\
    \midrule
    HDFS        & 2000 & 14 & 0.34 \\ 
    Hadoop      & 2000 & 114 & 0.74 \\ 
    BGL         & 2000 & 120 & 0.24 \\ 
    Thunderbird & 2000 & 149 & 0.48 \\
    \bottomrule
    \end{tabular}
\end{table*}

In the end, it is worth mentioning that there might be many events that are not relevant for anomaly detection. Such events are likely to be incorrectly clustered as, from our model's point of view, they do not contain useful information. 

\begin{figure*}[t]
  \centering
  \includegraphics[width=\columnwidth]{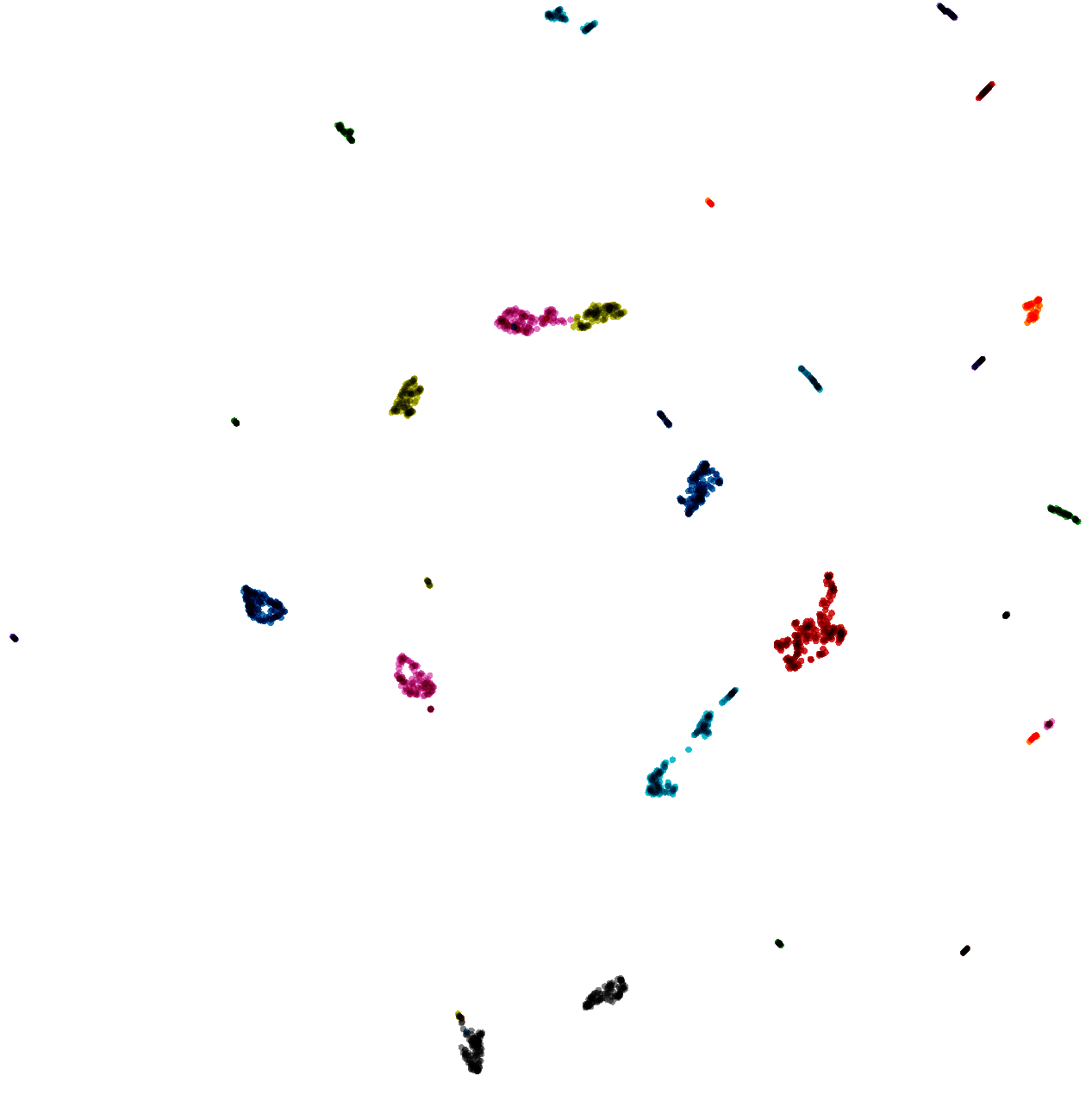}
  \caption{The embedded vectors visualization of the HDFS log parsing benchmark dataset.}
  \label{fig:embeddings}
\end{figure*}

Figure \ref{fig:embeddings} shows our visualization. Event distances are the distances of embedded log event vectors after U-Map transformation. Event colors represent the ground truth of manual labeling. We can see that events of the same color are closer to each other compared to events of different colors. This gives visual support to the idea that the Parser model processed events have formed clusters based on log text structure within the embedding vector space. We can see some points overlapping with incorrect clusters. This finding aligns with the Silhouette score, indicating that the clusters are imperfect. However, as our model is not trained for event parsing or labeling, it is irrational to expect perfect clusters here. Instead, this finding demonstrates our model's capability to learn some parsing rules as part of its end goal of anomaly detection. 
\textbf{Note:} Although the visualization experiment is possible for the Hadoop, BGL, and Thunderbird datasets, we found their images uninterpretable. HDFS only contains less than 15 event types, while the number is much higher for Hadoop, BGL, and Thunderbird, resulting in indistinguishable colors in the visualization.

\subsubsection{Discussion} It appears the Parser model learns to parse as a side product of its end goal of anomaly detection. An interesting future research idea would be to use just the parser model and see how it performs if trained for parsing only. One needs to remember that in the current setup, the Parser model has not been trained with the ground truth. Rather its silhouette scores are a side product of learning how to detect anomalies.

\subsection{Classifier Model Interpretation}
\subsubsection{Background}
Classifier Model is the second part of the HierarchicalCNN model. The classifier model, comprised of the sequence embedding component and classifier, is responsible for feature extraction and classification of the embedded events. The input to this component is the embedded event vectors (from Parser-model), while the output is a binary value indicating the input's anomaly status.

\subsubsection{Method}
This experiment focuses on finding the responsible event(s) for labeling a sequence as anomalous. This task is similar to finding responsible pixels (or areas) for image classification. Hence, we decided to choose the popular integrated gradients method \citep{integratedgrad}, which has been employed for responsible pixel task regularly. However, our entries are embedded events rather than pixels. Thus, we generated many heatmaps using the integrated gradient algorithm.

\subsubsection{Results}
\begin{figure*}
    \includegraphics[width=\linewidth]{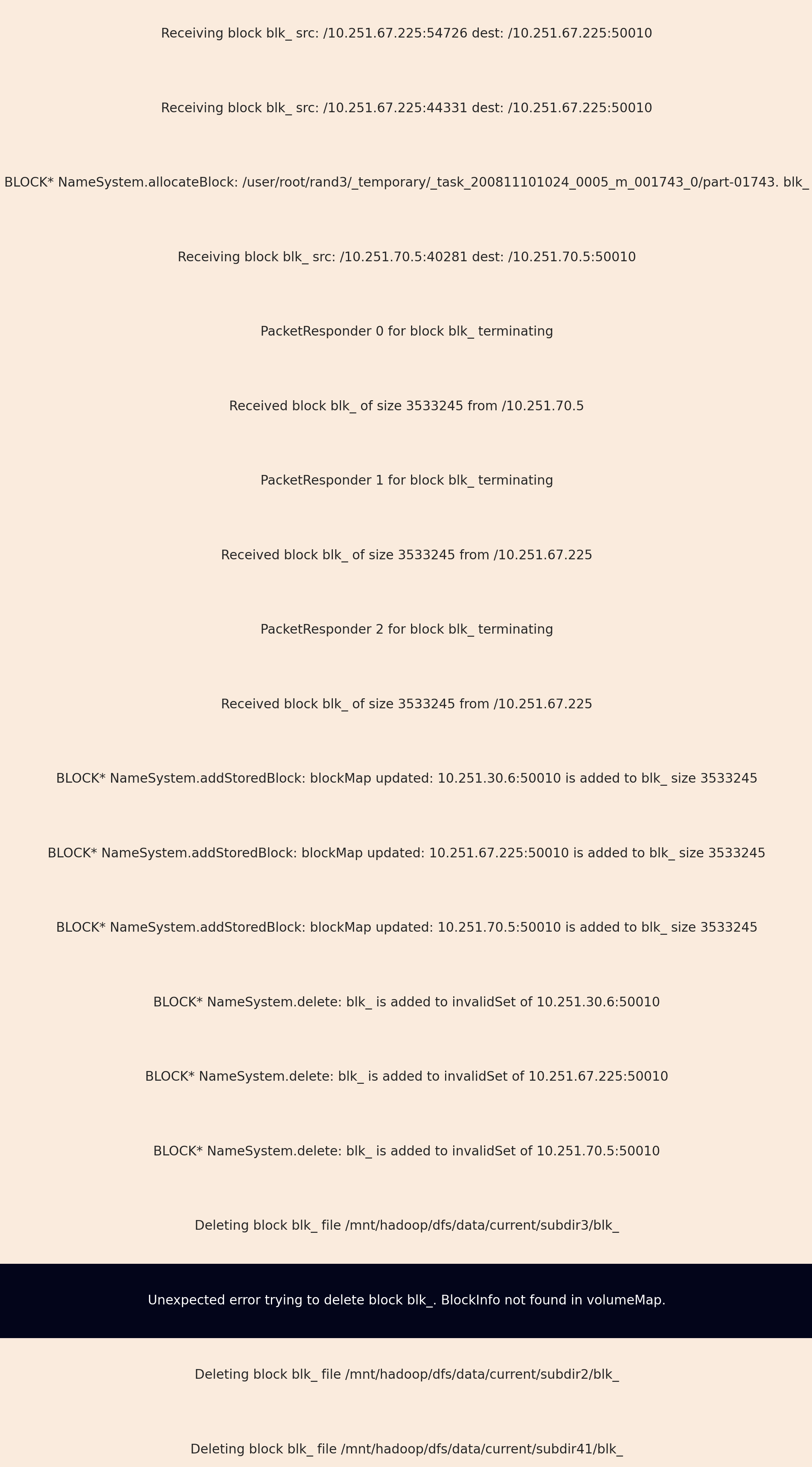}

    \caption{Suspiciousness map of various anomalous samples from the HDFS dataset that were detected as anomalies by the model for the occurrence of a fatal event. Darker colors represent more suspicious events.}
    \label{fig:fatal}
\end{figure*}

\begin{figure*}
    \includegraphics[width=\linewidth]{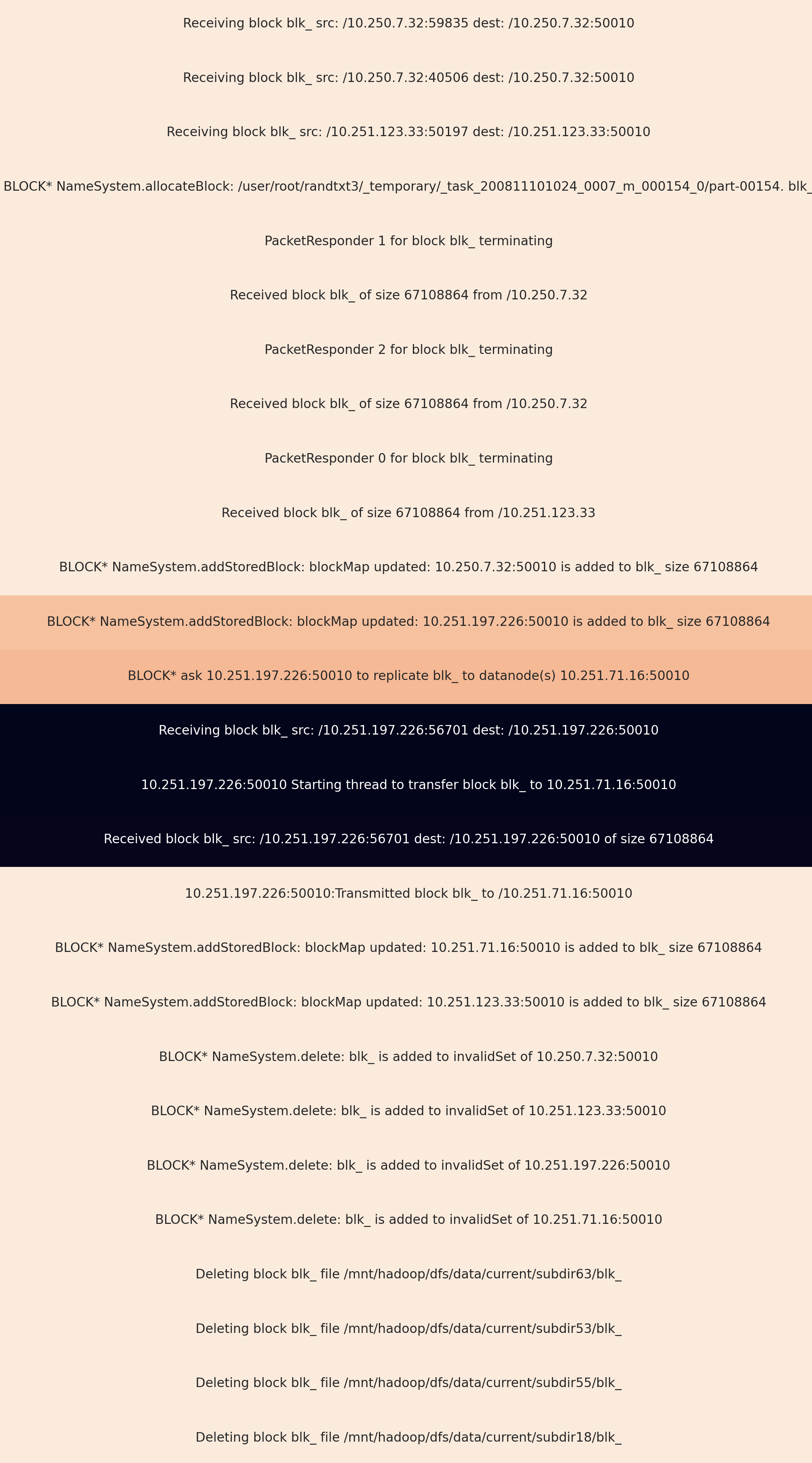}
    \caption{Suspiciousness map of various anomalous samples from the HDFS dataset that were detected as anomalies by the model for the occurrence of a suspicious sequence. Darker colors represent more suspicious events.}
    \label{fig:sequence}
\end{figure*}

\begin{figure*}
    \includegraphics[width=\linewidth]{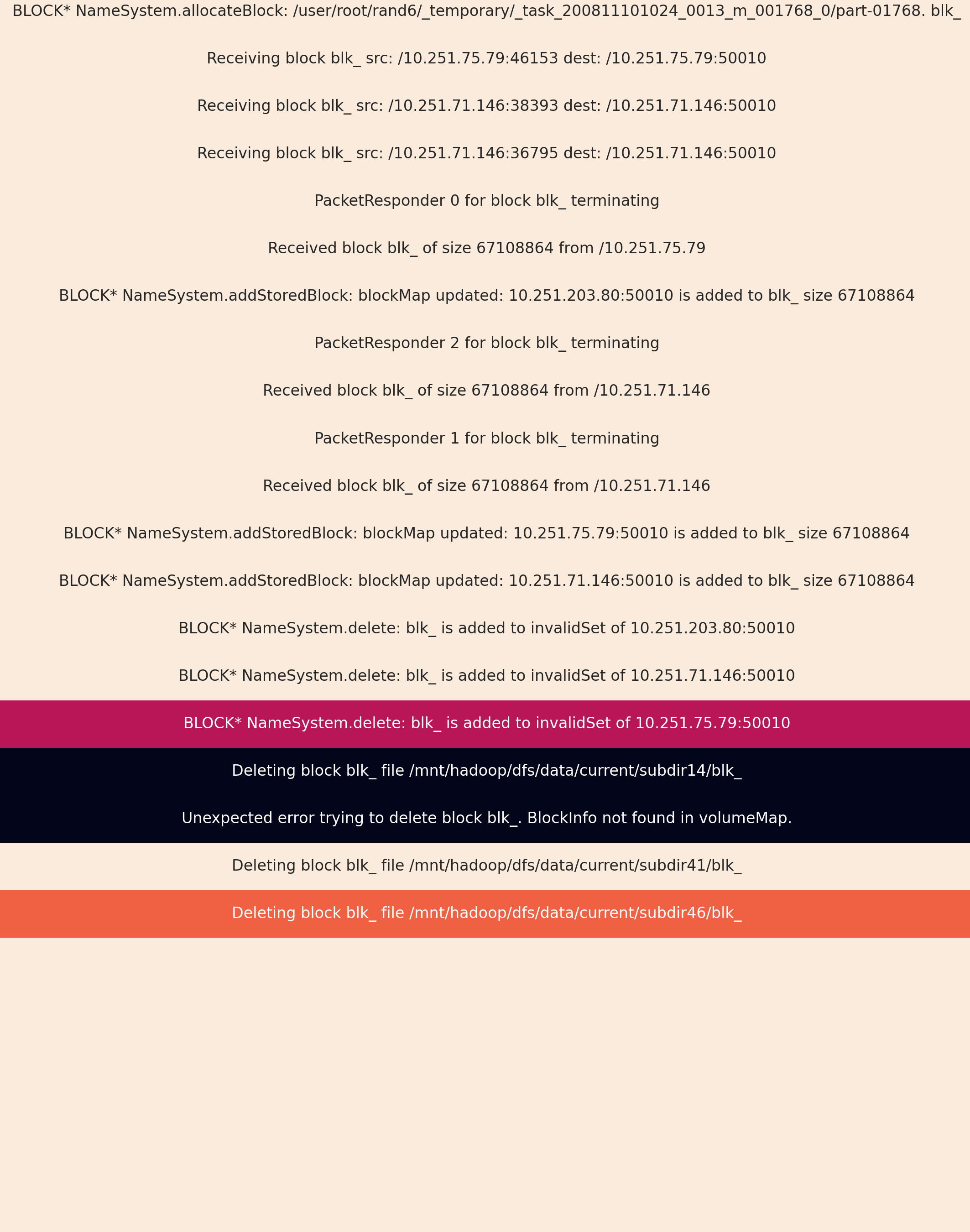}
  \caption{Suspiciousness map of various anomalous samples from the HDFS dataset that were detected as anomalies by the model for the occurrence of a suspicious event or sequence and a cleanup operation at the end. Darker colors represent more suspicious events.}
  \label{fig:cleanup}
\end{figure*}

\Cref{fig:fatal,fig:sequence,fig:cleanup} show selected anomalous sequence heatmaps (darker colors represent greater integrated gradient values).

We qualitatively explored the results for the HDFS dataset, shown in \Cref{fig:fatal,fig:sequence,fig:cleanup}, as this dataset is the only one with not many event templates and is comprehensible for humans. Furthermore, we found that the model relies on multiple strategies to classify anomalous sequences. According to the experiment, the following strategies explain what we think are three of the most important ones that the model utilized to classify anomalies and are relatively interpretable for humans.

\begin{enumerate}
    \item Fatal event: Detecting a single fatal event in the sequence, as shown in Figure \ref{fig:fatal}. This strategy is very straightforward as some events in the HDFS dataset occur only in anomalous situations, explained by \cite{sialog}.
    \item Bad subsequence: Detecting faulty subsequences. On this occasion, each sequence event does not indicate anomalous behavior individually. Nevertheless, the occurrence of them together in a particular order indicates that the entire sequence is anomalous, as examples in Figure \ref{fig:sequence}.
    \item  Multiple suspicious events: The last strategy exists when multiple suspicious events happen during the sequence, which may not indicate anomalous behavior on its own. For example, the software starts to clean up at the end. The model considers the combination of the suspicious event(s) and cleanup process as an anomalous action; see Figure \ref{fig:cleanup}.
\end{enumerate}


\begin{figure*}
    \includegraphics[width=\linewidth]{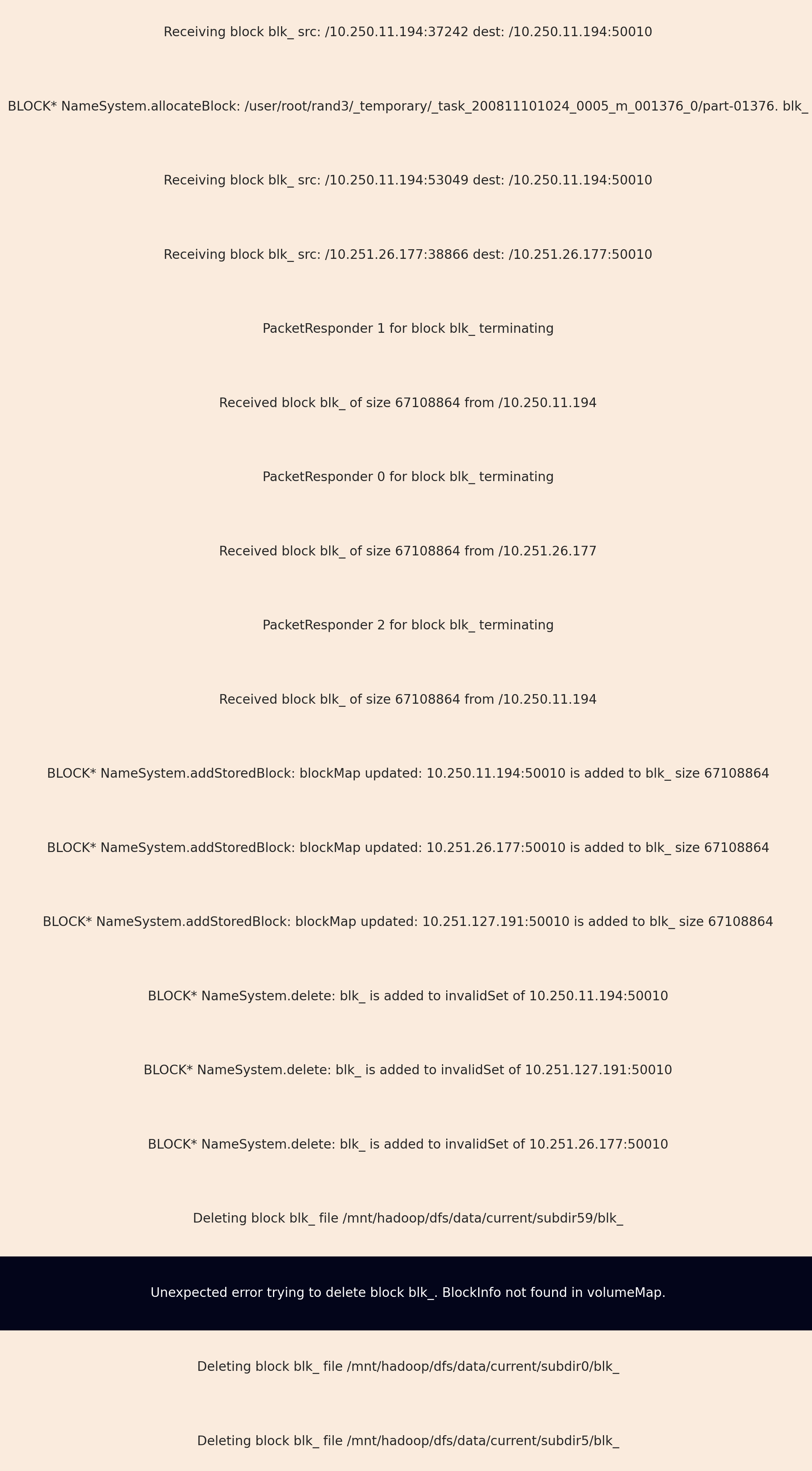}
  \end{figure*}
  \begin{figure*}
    \includegraphics[width=\linewidth]{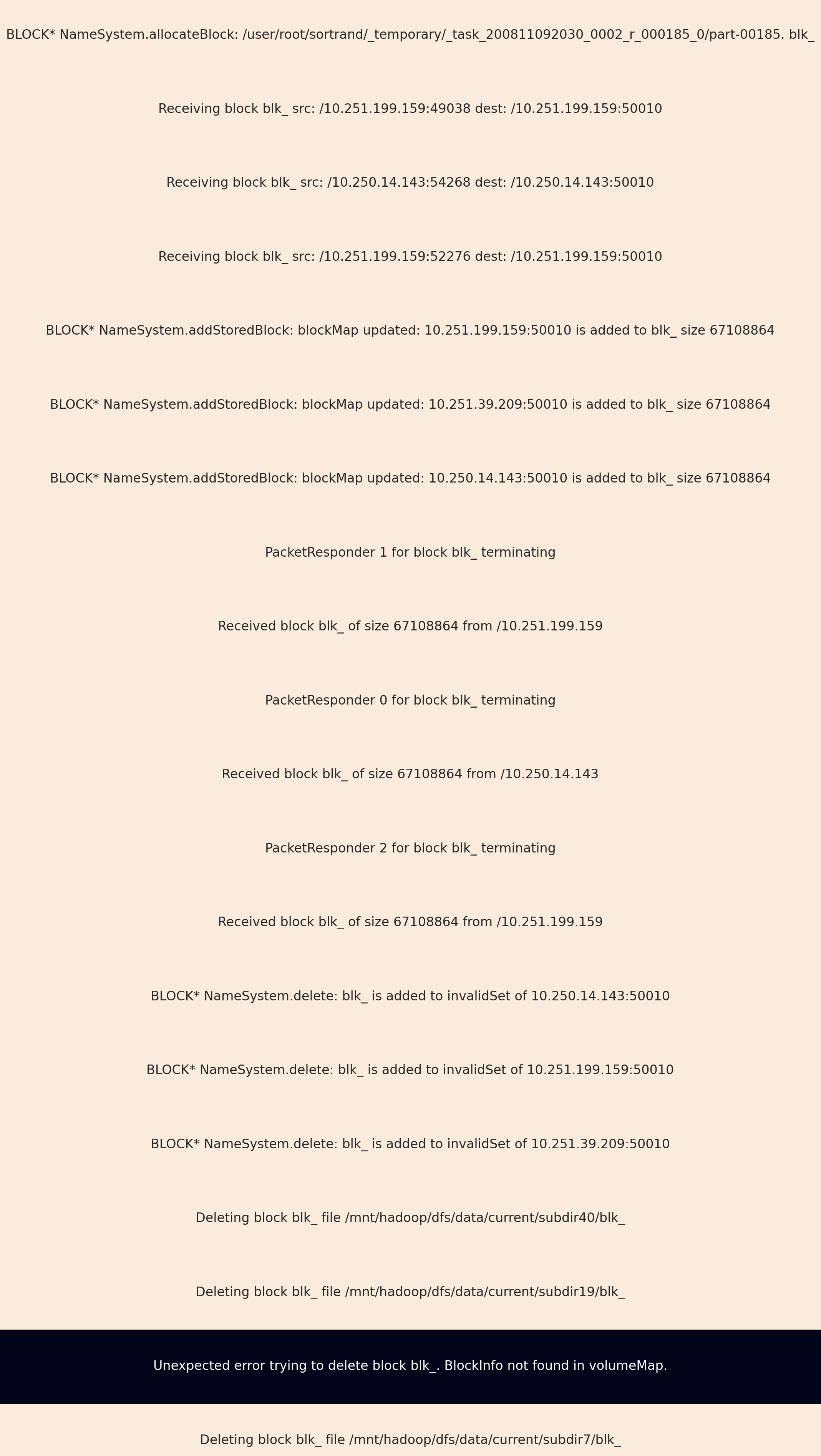}
    \end{figure*}

\begin{figure*}
    \includegraphics[width=\linewidth]{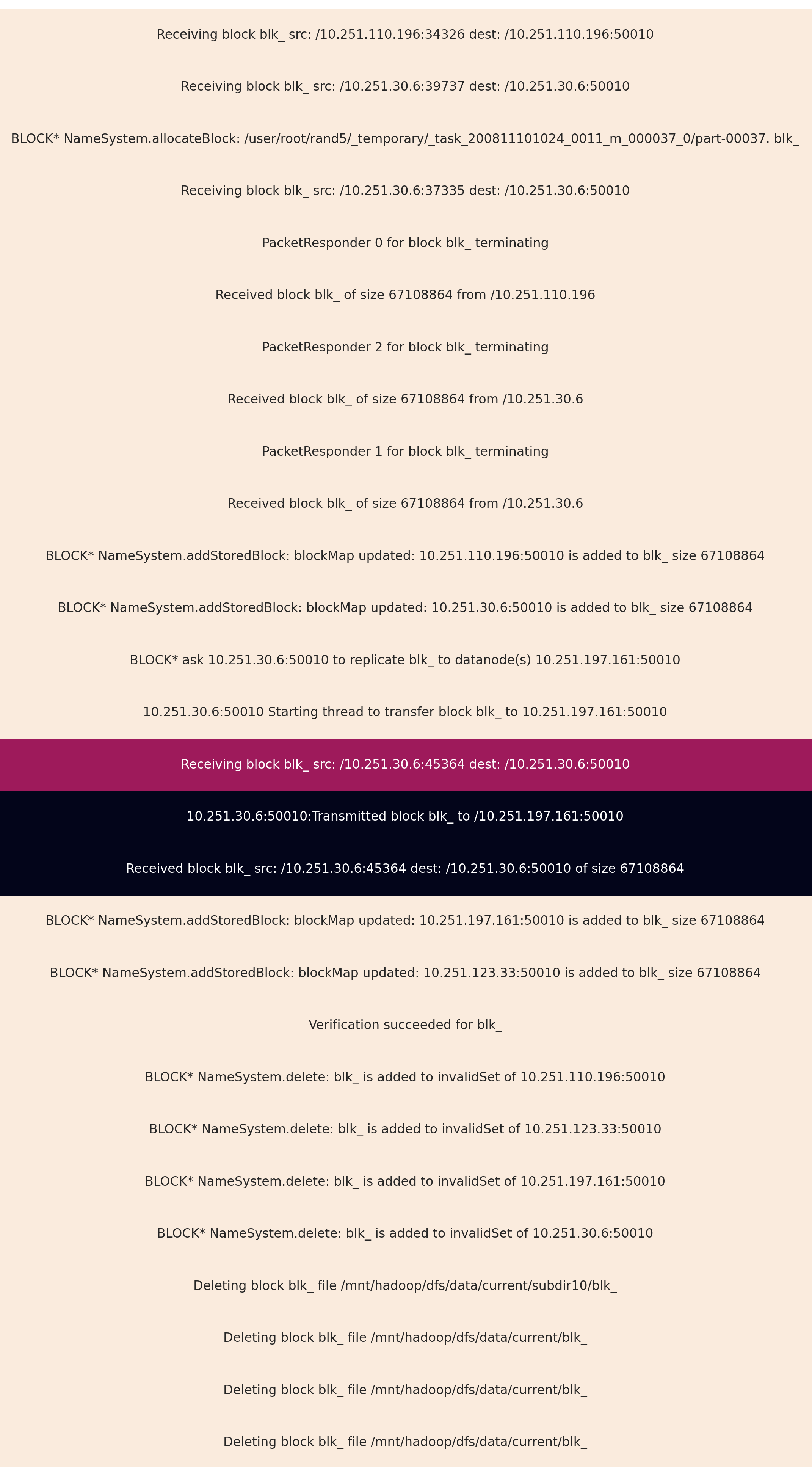}
  \end{figure*}
  \begin{figure*}
    \includegraphics[width=\linewidth]{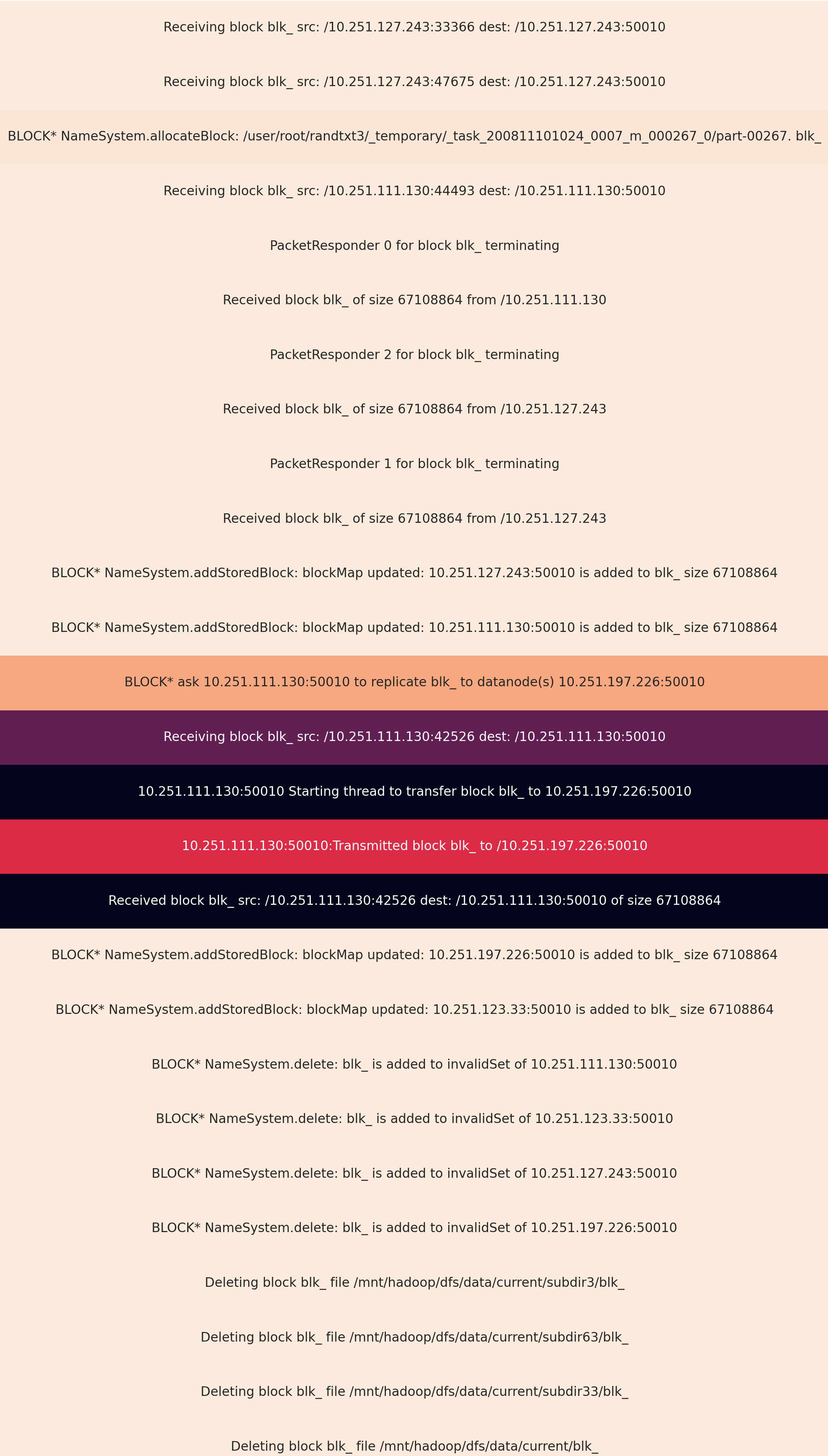}
  \end{figure*}

\begin{figure*}
    \includegraphics[width=\linewidth]{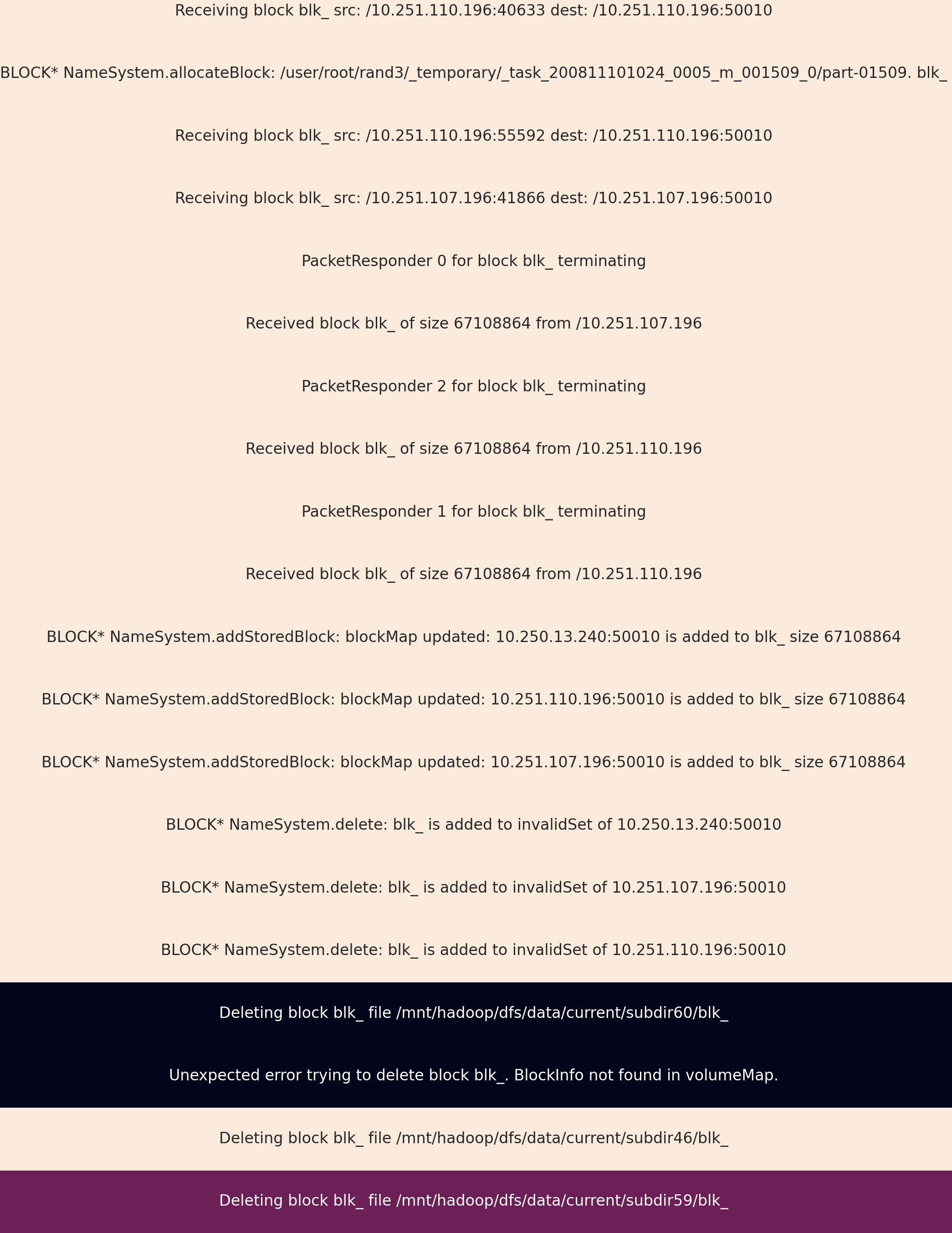}
  \end{figure*}
  \begin{figure*}
    \includegraphics[width=\linewidth]{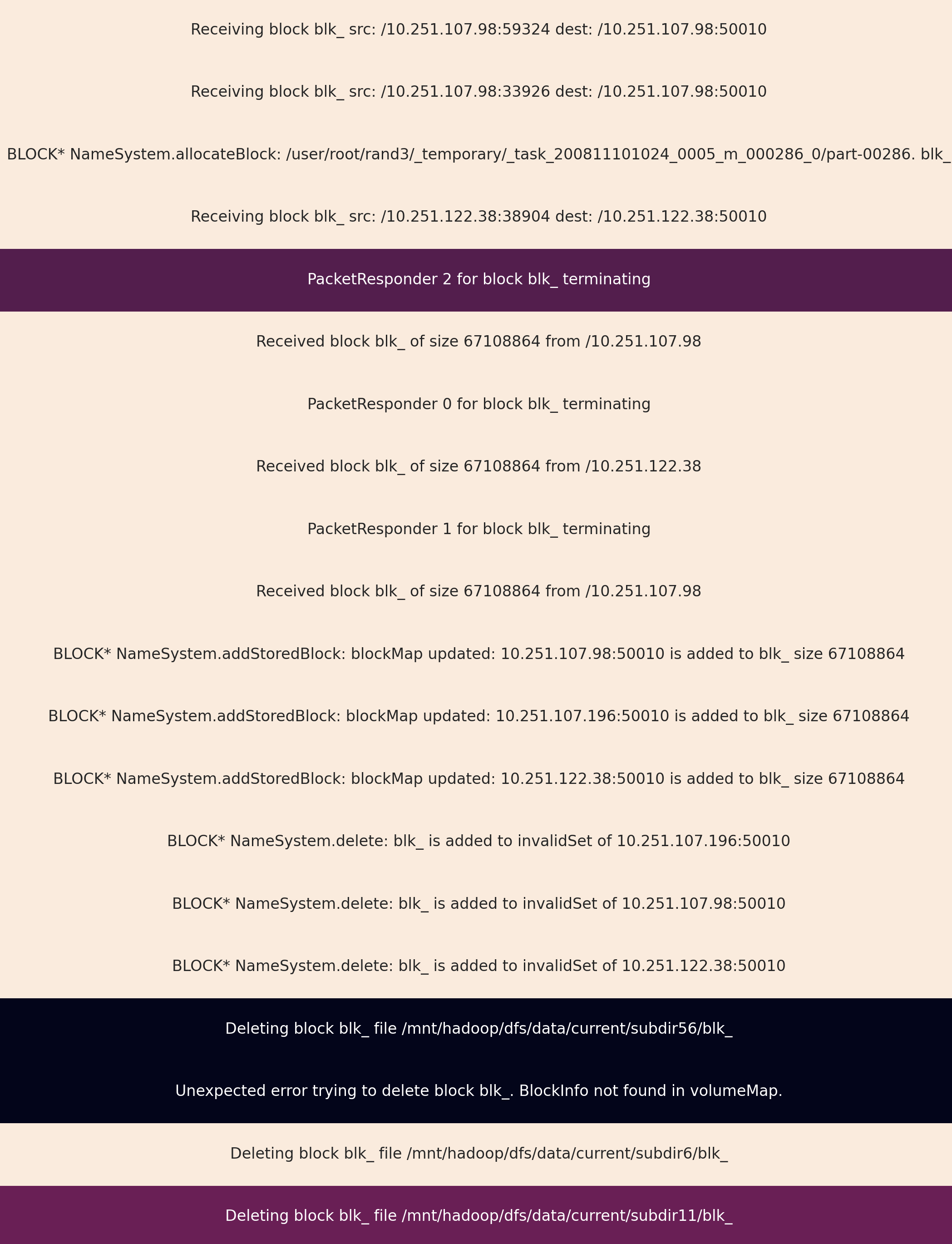}
  \end{figure*}


\section{Related works}
\label{sec:rel}

\begin{table*}
  \caption{Previous methods explanation and comparison to OneLog. Abbreviations include SM: Sequence Modeling, BC: Binary Classification, SV: Semantic Vectorization, TA: Template Approximation, T2V: Template2Vec, ML: Metric Learning, and HCNN: Hierarchical Convolutional Neural Network}
  \label{tab:prevmethods}
  \begin{center}

\begin{tabular}{llllll}
\toprule
Method     & Parser & Vectorizer & Classifier & Approach & Input Type    \\
\midrule
DeepLog\citep{deeplog}     & Spell  & Onehot & LSTM   & SM      & Sequence  \\ 
CNNLog\citep{cnnlog}     & -     & Onehot & CNN  & BC & Sequence \\ 
LogRobust\citep{logrobust}  & Drain & SV & Bi-LSTM & BC & Sequence \\ 
LogAnomaly\citep{loganomaly} & TA & T2V & LSTM & SM & Sequence \\ 
Logsy\citep{logsy} & -  & Tokenizer  & Transformer & BC & Message  \\ 
NeuralLog\citep{anodetnoparsing} & - & Tokenizer & Transformer & BC & Sequence  \\ 
Logbert\citep{logbert} & Tokenizer & - & Transformer & SM & Message  \\ 
Auto-BLSTM\citep{autoblstm} & Tokenizer& Autoencoder & Bi-LSTM & BC & Message  \\ 
SiaLog\citep{sialog}     & Drain  & Onehot & LSTM & ML & Sequence \\ 
\textbf{OneLog}     & \textbf{HCNN} & \textbf{HCNN} & \textbf{HCNN} & \textbf{BC} & \textbf{Both}\\
\bottomrule
\end{tabular}
\end{center}
\end{table*}

Here we compare our approaches to related work in terms of model design and performance. Although different approaches have been invented to address log anomaly detection, we focus on deep-learning-based models only as they have achieved the best performance in almost every dataset. 

Table \ref{tab:prevmethods} summarizes the prior works mentioned in this section. In the table, we can see that there are other approaches that do not utilize pre-built log parsers like Drain or Spell or works that have found innovative ways to retain semantic information available in log messages. Recently many works have utilized word tokenizer followed by Transformer, e.g., \cite{logsy,anodetnoparsing, logbert}. Yet, they all convert the log message input to words while we take the log messages as raw a character stream. Therefore, the most notable finding from the table is that prior works have used separate components for Parser and Vectorizer, while we have a single deep model with a Hierarchical Convolutional Neural Network (HCNN).
 
As one of the first deep-learning-based approaches, DeepLog \citep{deeplog} parses the sequences using the Spell log parsing algorithm mentioned by \cite{spell} and uses Long Short-Term Memory (LSTM) \citep{lstm} to model the non-anomaly logs by predicting the next event in the sequence. After the training, the model predicts a low probability for anomaly sequences as it has converged on non-anomaly data. DeepLog also uses parameter values and log keys to preserve text information.  

LogRobust \citep{logrobust} applies an attention model to Bidirectional Long Short-Term Memory (Bi-LSTM) to classify the event sequences. However, the main contribution of this work is in the introduction of log evolution. LogRobust asserts that software logs evolve due to updates. Hence, it proposes a method for synthetically evolving log messages and sequences by adding noise. Furthermore, it introduces a new method of vectorization called ``semantic vectorization'', which uses pre-trained word embeddings to construct a vector based on words' semantic meanings in a log event. In a way, this semantic vector can be seen to retain some natural language information.

LogAnomaly \citep{loganomaly} also uses LSTM but presents a new vectorization technique called ``template2vec' that takes advantage of synonyms and antonyms. This template approach is somewhat similar to LogRobust, where part of natural language information is retained through the words in the template. Additionally, LogAnomaly keeps track of message counts in sequences to detect quantitive anomalies alongside sequential ones. 

In another innovation in the vectorizer component, \cite{cnnlog} proposes a technique to embed log keys to feature-rich vectors called ``log-key2vec''. Besides, it uses Convolutional Neural Network (CNN) \citep{cnn} to classify sequences, making the model more computationally efficient in training and inference time compared to LSTM-based models. A parser is also used in this work, yet, the details are unclear. 

The first work to propose the adoption of Transformer \citep{bert} in log anomaly detection is Logsy \citep{logsy}. Logsy embeds log messages in a vector space, with non-anomaly messages clustered at the origin while anomaly messages embedding at a distance. A unique loss function, which enables the learning process of embedding operations, is also among its contributions. The second  case of using the Transformer, NeuralLog, mentioned in \cite{anodetnoparsing}, uses a Transformer Encoder on top of a pre-trained Bert model to take advantage of both the semantic embedding of Bert and the self-attention mechanism of the Transformer Encoder. \cite{anodetnoparsing} achieves relatively high scores among multiple datasets. The third approach is using Transformer \cite{logbert} uses Bidirectional Encoder Representation from Transformer (BERT) to learn normal data patterns and use it to identify anomalies in a semi-supervised fashion.

In another unique approach, \cite{autoblstm} introduces Auto-LSTM, Auto-BLSTM, and Auto-GRU, which also operate on natural language directly without needing a parser. They first extract features from log messages using an autoencoder \citep{autoencoder}. Then a recurrent neural network module, namely LSTM, Bidirectional LSTM, or GRU, is used to classify the message. The method achieves accurate results due to its nice pipeline of neural networks. 

Finally, in our past work \citep{sialog}, we utilized the Siamese network \citep{siamese} with LSTM to embed log sequences into a vector space that keeps sequences of the same type (anomaly / non-anomaly) close to each other while maximizing distance from the different type. Additionally, the authors introduce other benefits that come with the Siamese network, such as more robust predictions, unsupervised evolution monitoring, and sequence visualization.

\section{Threads to Validity} 
\label{sec:lim}

Despite OneLog's superior performance over other software log anomaly detection methods within a multi-project framework, it is not devoid of limitations. The most substantial challenge lies in the requirement for labeled data, which can be a significant obstacle in many practical scenarios. Additionally, the computational expense associated with the deep neural network presents another potential limitation. In the subsequent subsections, we will thoroughly examine our research's internal and external validity. This exploration will encompass a detailed discussion to assess the robustness and applicability of our findings across various contexts.

\subsection{Internal Validity}
Regarding internal validity, it is imperative to note that the evaluation environment of OneLog diverges significantly from alternative methodologies due to their mutual incompatibility. In methodologies necessitating parsing, the parser's elimination of parameters from log messages results in the homogenization of numerous log messages. This leads to substantial duplication within the dataset, necessitating a duplication removal prior to splitting the dataset into training and testing subsets. Consequently, achieving identical evaluation environments is infeasible.

In order to circumvent this problem, we performed the experiments multiple times with different random seeds and ensured we were getting consistent results. Furthermore, we reimplemented state-of-the-art methods (except ones marked with a superscript asterisk in Table \ref{tab:singleresults}) and ensured that the outcomes aligned with, if not surpassed, the original results. It is noteworthy to mention that certain methodologies, notably DeepLog, exhibited enhanced performance in our implementation relative to their initial publication.

All in all, We are confident in our work since we evaluated it in a variety of contexts, including different datasets and model capabilities. The fact that we achieved consistent scores in diverse circumstances proves that our suggested strategy works as predicted and leads to more accurate software log anomaly identification.

\subsection{External Validity}
Concerning the matter of external validity, it is pertinent to acknowledge that it may be subject to limitations due to the unavailability of an appropriate industrial dataset. This absence significantly constrains our capacity to assert the efficacy of our methodology within a real-world industrial setting. However, it is essential to note that the primary objective of this study was to introduce a novel end-to-end anomaly detection methodology and conduct its preliminary validation. Consequently, a pressing need exists for subsequent research endeavors to thoroughly explore the advantages and limitations of OneLog and Hierarchical Convolutional Neural Network architectures in the context of software log anomaly detection in real-world settings.

\subsection{Construct Validity}
Construct validity in software engineering involves ensuring that the measurements, tests, or procedures used in a study actually measure the theoretical constructs they are intended to measure \cite{wohlin2012experimentation}. We used well-established statistics from machine learning like $F_1$ score, so we see no problem there. However, regarding our data there is of course the issue that labeled anomalies might not represent true or anomalies that would be important enough to software operations engineers. There is very little we can do to mitigate this issue.

\section{Implications}
Character-Based Approach and Parser Elimination:
Our design employs a character-based processing method for log events, enhancing performance by leveraging elements such as numbers and punctuation, which are often disregarded in word-based or parser-based approaches. By incorporating complete messages, OneLog is adept at detecting anomalies that may arise from incorrect parameters. This raises an intriguing possibility: end-to-end learning systems like OneLog might obviate the need for traditional log parsers, a topic that has garnered considerable research interest. This shift could streamline log analysis processes, making them more efficient and effective.

Use of Multiple Datasets at Once:
Our approach facilitates the simultaneous utilization of multiple datasets which is currently supported in one prior work \citep{logsy}. This feature is particularly beneficial in scenarios where available training data are limited, allowing for the augmentation of smaller datasets with a broader body of public datasets. Our findings demonstrate that this strategy substantially enhances model performance. Moreover, we have successfully implemented cross-project anomaly detection, leveraging only external data sources, which is effective when analyzing system logs of sufficient similarity. These steps mark an advancement towards the integration of transfer and augmented learning principles within the realm of log anomaly detection, underscoring OneLog's potential to overcome traditional data scarcity challenges.

Model Interpretations Enhance Confidence in OneLog:
Through model interpretations, we have uncovered that OneLog autonomously develops human-like event parsing rules. Beyond parsing, OneLog utilizes at least three distinct anomaly detection rules: fatal events, bad subsequences, and multiple suspicious events. Our analysis of the model's internal logic affirms our confidence in OneLog's capabilities, showcasing its potential as a solution in the field of log analysis.

\section{Conclusion and Future Works}
\label{sec:con}

This paper presents OneLog, a novel method to detect anomalies in software logs. OneLog merges the parser, vectorizer, and classifier components into one deep neural network, which moves the log analysis field closer to complete end-to-end learning. We think the most interesting findings are. First, producing state-of-the-art performance; Second, training the model on sequence-based and event-based datasets with no modification; Third, making multiple datasets usage possible in training to enhance performance; Fourth, model interpretations showing that OneLog internally learns human-like event parsing rules and anomaly detection rules.

Although the model results trained on the multi-project dataset are good, it does not come without any room for improvement. As a matter of fact, the lack of sufficient data volume might result in poor performance, as observed in some of our experiments. Therefore, we believe more extensive datasets would contribute to this work in future works. Furthermore, additional datasets from different projects or even the same project with varying versions may also benefit the multi-project dataset. Conversely, if OneLog could be systematically architected to utilize normal logs (non-anomalous) as the primary training source, the data acquisition challenge might be substantially mitigated, since, normal data, in contrast to anomalous data, is more readily produced and accessible, offering OneLog a more sustainable and efficient solution for training.

Finally, the goal of the paper was to develop end-to-end learning in log analysis, yet as shown in Figure \ref{fig:teaser}, we still have the preprocessor stage. So how could we remove it? Our preprocessor is simple as it only organizes the raw data into sequences. Hence, to improve the end-to-end learning further, we could have a more generalizable sequence creation method for all datasets. This would facilitate the process of using multiple datasets without any human intervention.

\section{Acknowledgment}
This work has been supported by the Academy of Finland (grant IDs 298020 and 328058). Additionally, the authors gratefully acknowledge CSC – IT Center for Science, Finland, for their generous computational resources.



\bibliographystyle{plainnat}
\bibliography{refs}
\end{document}